\documentclass{optica-article}

\journal{opticajournal} % for journals or Optica Open

\articletype{Research Article}

\usepackage{listings}
\usepackage{subcaption}
\usepackage{graphicx}
\usepackage{floatrow}
\usepackage{verbatim}

\DeclareRobustCommand{\uvec}[1]{{%
	\ifcsname uvec#1\endcsname
		\csname uvec#1\endcsname
	\else
		\bm{\hat{\mathbf{#1}}}%
	\fi
}}

\newcommand{\imi}{\mathrm{i}}
\newcommand{\eue}{\mathrm{e}}

\usepackage{libertinus}

\begin{document}

\title{Multi-mode Heterodyne Laser Interferometry Realized via Software Defined Radio}
%\title{Multi-mode Heterodyne Detection Scheme Realized via Software Defined Radio for Sensitivity-Enhanced Laser Interferometry}

\author{X. Lin\authormark{1}, M. T. Hartman\authormark{1}, S. Zhang\authormark{1,2}, S. Seidelin\authormark{3}, B. Fang\authormark{1} and Y. Le Coq,\authormark{1,*}}

\address{\authormark{1}LNE-SYRTE, Observatoire de Paris, Universit\' e PSL, CNRS, Sorbonne Universit\' e, Paris, France\\
\authormark{2}currently with Menlo Systems GmbH, Bunsenstraße 5, D-82152 Martinsried, Germany\\
\authormark{3}Univ. Grenoble Alpes, CNRS, Grenoble INP and Institut N\' eel, 38000 Grenoble, France}

\email{\authormark{*}yann.lecoq@obspm.fr} %% email address is required; see note below about the corresponding author designation

% use {asbstract*} to suppress the copyright line. Copyright information will be added in production

\begin{abstract*} 
The agile generation and control of multiple optical frequency modes combined with the realtime processing of multi-mode data provides access to experimentation in domains such as optomechanical systems, optical information processing, and multi-mode spectroscopy.  The latter, specifically spectroscopy of spectral-hole burning (SHB), has motivated our development of a multi-mode heterodyne laser interferometric scheme centered around a software-defined radio platform for signal generation and processing, with development in an entirely open-source environment.  A challenge to SHB is the high level of shot noise due to the laser power constraint imposed by the spectroscopic sample.  Here, we have demonstrated the production, detection, and separation of multiple optical frequency modes to the benefit of optical environment sensing for realtime phase noise subtraction as well as shot noise reduction through multi-mode averaging.  This has allowed us to achieve improved noise performance in low-optical-power interferometry. Although our target application is laser stabilization via SHB in cryogenic temperature rare-earth doped crystals, these techniques may be employed in a variety of different contexts.

%The ability to precisely prepare, control and probe a physical state is at the center of a variety of experiments exploiting the classical and quantum properties of dopants in a crystal. For instance, applications such as frequency-locking a laser to a spectral hole, or creating optical classical- and quantum memories can all be achieved using for instance rare-earth doped crystals, as long as the physical states of the dopants are highly controlled. Often, such manipulations require the ability to address several frequency channels in parallel, which in turn relies on signal demodulation, and the generation of frequency corrections. We here show that this can been readily achieved with a hardware and software implementation based on off-the-shelf commercially pre-programmed field-programmable gate array and a Software Defined Radio platform. \textcolor{red}{YLC copy pasted here the previous abstract. need substantial re-writting for OE version}

\end{abstract*}

%%%%%%%%%%%%%%%%%%%%%%%%%%  body  %%%%%%%%%%%%%%%%%%%%%%%%%%
\section{Introduction}
Optical heterodyne detection is a well developed technique, having seen innumerable applications in science and industry, where the measurement of a relatively weak optical signal is achieved by interference with a relatively strong local oscillator to produce a detectable beat note.  Implementations of the technique vary widely through the domains of industrial application, including fiber communication \cite{1075396}, LiDAR and atmospheric sciences \cite{TheLaseranditsApplicationtoMeteorology}, astronomy such as radiometry \cite{McElroy:72} and astrophysical spectrometry \cite{Abbas:76}, instrumentation and metrology such as laser frequency stabilization \cite{PhysRevD.92.022004, Gobron:17, Galland:20}, and fundamental physics, for example gravitational-wave detection \cite{Abbott_2009} and the search for axion-like particles \cite{DIAZORTIZ2022100968, HALLAL2022100914}, just to name a few.

In many use cases of spectroscopy the probe beam is constrained to low optical powers, either by the sources including astronomical applications or to avoid damaging the sample, for instance in many biological applications \cite{C8RA04491K, molecules27196752}.  Spectral hole burning (SHB) is a spectroscopic technique of imprinting spectral lines in an inhomogeneously-broadened absorption spectrum \cite{Hercher:67}.  This is accomplished by using a pump-laser to selectively excite an absorptive dye or dopant atoms/ions in the host to a higher state, and, upon returning to the ground state, they occupy a different hyperfine level which is off-resonant with pump light.  The result is a narrow transmissive spectral line at the pump-laser frequency that has been `burned' into the absorption spectrum.  When probing these imprinted spectral features, the laser power used is constrained due to the risk of `overburning', where the use of relatively high power causes, after some time, a broadening of the spectral hole leading to a degradation in the spectroscopic sample \cite{MANILOFF1995173}.

Rare-earth ion doped crystals present an attractive medium for spectral-hole spectroscopy by providing optical transitions with large coherence times (order ms) and hyperfine splitting at the ground state which provide long-lived population states at cryogenic temperatures.  Spectral-hole spectroscopy of rare-earth ions can be, and has been, used in many applications ranging from classical~\cite{Berger:16, Venet:18} and quantum~\cite{NILSSON2005393, Bussieres2014, walther2015:PhysRevA.92.022319, Maring2017} information processing to optomechanics~\cite{Molmer2016:PhysRevA.94.053804, seidelin2019:PhysRevA.100.013828} and high-precision laser measurements and stabilization~\cite{Julsgaard:07, PhysRevLett.107.223202, Thorpe2011, Thorpe_2013}. The domain where our experimental group concerns itself is a metrology-purposed SHB heterodyne spectroscopy, namely its application to the stabilization of laser frequency to spectral holes burned into a rare-earth ion doped crystal, $\mathrm{Eu}^{3+}:\mathrm{Y}_{2}\mathrm{Si}\mathrm{O}_{5}$.  Each of these applications requires precise and agile manipulation of laser frequency for the burning and probing of spectral holes.  For our purpose, we developed a laser control system based on a Software Defined Radio (SDR) platform which provides a flexible, easily customizable, and greatly extensible field-programmable gate array (FPGA)-based tool for acquisition and control.  We have implemented this platform successfully in our experiment, demonstrating its use in laser stabilization to, as well as physical characterization of, spectral holes \cite{Gobron:17, Galland:20, Galland2020:PhysRevApplied.13.044022, Zhang2020aPhysRevResearch.2.013306, Zhang2020:doi:10.1063/5.0025356, Zhang2023:PhysRevA.107.013518} .  

In the context of laser frequency stabilization via SHB, the laser power constraint (of order $10\,\mathrm{nW}$ for continuous probing of a single spectral hole) to avoid hole overburning leads to a substantial level of shot noise in the heterodyne probe signal resulting in laser frequency error, random in time, in the detection process. In this paper, we address this challenge by extending our SDR-based control system to produce, detect, and demodulate multiple heterodyne beat signals, and we demonstrate the subsequent reduced detection noise obtained by averaging over these modes.  This novel technique allows us to simultaneously probe several spectral holes and synthesize an error signal from any combination of resulting heterodyne spectroscopy signals.  In addition to a simple average of the narrow spectral hole signals, we demonstrate the production of additional monitor modes, outside the band around the narrow spectral holes, which can measure and subtract unwanted differential optical phase noise between a reference heterodyne path and the spectroscopic sample path in real time.  We believe this technique and platform will benefit not only our ultra-stable laser experiment, but any spectroscopy or heterodyne interferometry scheme where either agile frequency control is required, or the power per frequency-mode is constrained, or an appropriate use of additional probing modes may help reduce the effect of technical noise sources, including other laser stabilization experiments (for example, iodine-based laser stabilization \cite{philippe_frequency_2016}). This article is organized as follows: a first section describes the general optical apparatus that is used in conjunction with the SDR-based platform. In a second part, we describe the main building blocks of the platform and the way it can be used to perform various experimental tasks pertaining to spectral hole burning in rare-earth-doped crystals at cryogenic temperatures. In a third part, we present the results of sensing noise measurements exploiting the fixed, low-power multi-mode heterodyne detection technique using our SDR-based system. A last section discusses the advantages and the limitations of this platform and explores ways to go beyond what is currently achieved.

\section{Experimental Apparatus}

Past iterations of our apparatus have been described succinctly at its various development stages in our publications~\cite{Gobron:17,Galland:20, Galland2020:PhysRevApplied.13.044022}. Our experimental apparatus is primarily built for exploring the use of spectral hole burning in $\mathrm{Eu}^{3+}:\mathrm{Y}_{2}\mathrm{Si}\mathrm{O}_{5}$ in the realization of an ultra-stable laser of unprecedented short-term (near $1\,\mathrm{s}$) frequency stability~\cite{Galland:20}.  The ultimate goal of this experiment is to produce a continuous wave (CW) laser whose performance allows optical lattice clocks \cite{Nicholson2012:PhysRevLett.109.230801} to reach their fundamental frequency stability limit, namely the so-called quantum projection noise limit \cite{Itano1993:PhysRevA.47.3554}. While this limit can be breached by quantum-engineered protocols (spin-squeezing, NOON states\cite{N00NStates:doi:10.1126/science.1188172}...), those implementations likewise necessitate a probe laser with ultra-high short term frequency stability. A secondary target of our experiment is to perform high-precision spectroscopy of micro-fabricated mechanical resonators realized via rare-earth-ion-doped crystals, in order to probe and control their oscillations down to the quantum level~\cite{seidelin2019:PhysRevA.100.013828}. Both scientific targets utilize the same apparatus and SDR-based platform, only the rare-earth-ion-doped crystal is changed between the two applications: a bulk centimeter-scale crystal for the ultra-stable laser application in contrast to various micro-fabricated samples for the micro-resonator application. Here we give an overview of the current setup and summarize parts that are particularly important to understand the use of our SDR-based platform in the next sections.

\begin{figure}[t]
    \centering
    \includegraphics[width=0.98\linewidth]{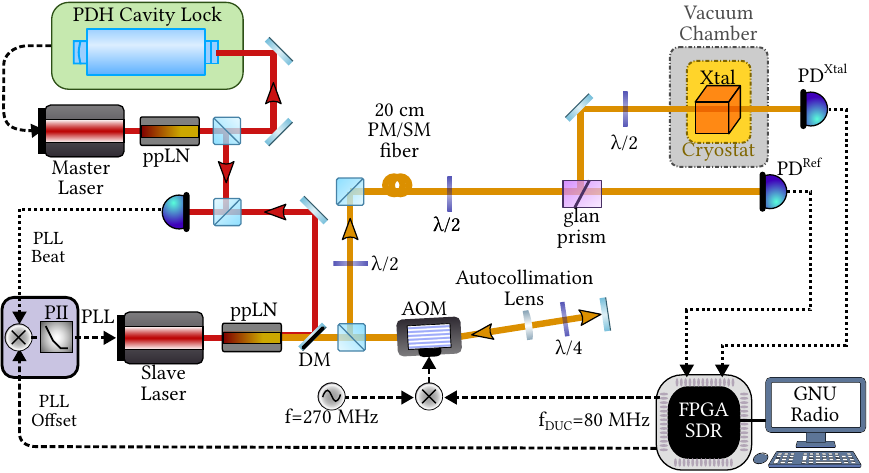}
    \caption{\label{fig:scheme} Illustration of our current experimental setup. ppLN =  periodically poled lithium niobate optical frequency doubler; PDH = Pound–Drever–Hall; DM = dichroic mirror; AOM = acousto-optic modulator; PLL = phase-lock-loop; PII = proportional+double integrator; $\mathrm{\lambda/2}$ = half-wave plate; PM/SM fiber = polarization-maintaining single-mode fiber; PD = photodetector.}
\end{figure}

An illustration of the experimental apparatus can be seen in Fig. \ref{fig:scheme}.  The Master and Slave lasers are extended cavity diode lasers (Toptica DL Pro) operating nominally at $1160\,\mathrm{nm}$.  The master laser, which is frequency locked via the Pound-Drever-Hall technique \cite{Drever1983, Black2001:10.1119/1.1286663} to a commercial metrological-grade Fabry-Perot cavity (Stable Laser Systems), serves only as a frequency reference and to provide an initial frequency stability to the slave laser, which is offset phase-locked to the master via a phase-lock loop (PLL). Adjusting the PLL offset allows us to coarse-tune the slaver laser frequency to the interesting spectroscopic sites which occur between the resonances of the reference cavity (Free Spectral Range $\approx 1.5\,\mathrm{GHz}$).  The slave laser, which is the main science laser for spectroscopy, is frequency doubled in fiber-coupled waveguide ppLN crystals (NTT Electronics) to the spectroscopic sites (nominally $580\,\mathrm{nm}$) of our typical sample, $\mathrm{Eu}^{3+}:\mathrm{Y}_{2}\mathrm{Si}\mathrm{O}_{5}$.

The yellow light of the slave laser is frequency shifted in a double-pass configuration \cite{Riza1996:10.1063/1.1147199} through a high-speed acousto-optic modulator (AA Opto-electronics, MT350-A0.12-VIS) to the exact desired spectroscopic frequencies.  The high-speed and large bandwidth of the acousto-optic modulator (AOM) are required to provide a high-fidelity reproduction of the wide range of frequency components in its radio-frequency (RF) drive signal.  It is at this point where we apply the fine frequency control in the generation of the multiple frequency modes in the slave laser beam, namely the spectroscopic probe modes, an optical path length monitor mode, and a local oscillator mode to beat the probe/monitor modes down to detectable RF heterodyne frequency.  The production of the AOM RF drive signal is realized through the output of the Ettus USRP X300 FPGA-Based Software Defined Radio (SDR).  The electrical RF signal containing the desired frequency modes is programmed via the GNU Radio environment in the controlling PC.  This process is explained in great detail in Section \ref{sec:SDR}.  The signal is Digitally Up-Converted (DUC) from base-band to the SDR output center frequency, $f_\mathrm{DUC}=80\,\mathrm{MHz}$, which is then up-mixed with a function generator (a $270\,\mathrm{MHz}$ sinusoid) to reach the center drive frequency of the AOM, $f_\mathrm{AOM}=350\,\mathrm{MHz}$, which is then amplified before driving the AOM.

The frequency shifted light is typically composed of several optical spectral modes with near identical spatial modes, thanks to the double-path AOM configuration. It is passed through a polarization-maintaining single-mode (PM/SM) fiber before being spatially separated into a reference path (`Ref', which provides the reference heterodyne beat note after free-space) and sample path (`Xtal', which, in spectral-hole experiments, contains the sample crystal housed in a cryostat (MyCryoFirm Optidry) within a vacuum system).  In laser stabilization experiments, the desired signal is an optical-frequency-dependant phase-shift, $\phi_{\mathrm{s}}$, of the optical mode along the Xtal path which, due to the Kramers-Kronig relation, is proportional to the Hilbert transform of the absorption in the crystal. When the mode is close to the center of a spectral hole, this phase shift is simply proportional to the frequency error, $\nu_\mathrm{err}\,[\mathrm{Hz}]$, between the laser probe-mode frequency and the spectral hole center frequency with the proportionality coefficient defining the frequency discriminant of the spectral hole, $D_{\phi/\nu}\,[\mathrm{rad}\,\mathrm{Hz}^{-1}]$,
\begin{equation}
    \phi_{\mathrm{s}} = \nu_\mathrm{err} D_{\phi/\nu}. \label{eqn:SpectroscopicSignal}
\end{equation}

The reference and signal beat-notes are detected at the $\mathrm{PD}^\mathrm{Ref}$ and $\mathrm{PD}^\mathrm{Xtal}$ avalanche photodetectors (Thorlabs, APD410A/M), respectively.  The beat signals are digitally down converted in the SDR and the signal phase is recovered via digitally extracting the phases of the various spectral components and comparing these quantities between the Xtal path and the Ref path within the computer control program. The clocks for the analog-to-digital converters (ADCs) (ie. the clock of the SDR) and the clock in the external upconverting function generators are all synchronized to a $10\,\mathrm{MHz}$ lab-wide distribution of a reference clock signal.  In a laser stabilization experiment, this real-time complex data treatment provides the error signal which acts on the slave laser PLL frequency offset actuating the slave laser towards the spectral hole center frequency. In this paper we demonstrate a multi-mode heterodyne interferometric readout where the simultaneous acquisition and real-time analysis of several optical frequency channels in the SDR platform is detailed in Section \ref{sec:SDR}. The detection sensitivity measurement results are discussed in Section \ref{sec:Multi-ch_lp_hetdet}.

\section{The Software Defined Radio Based Platform}
\label{sec:SDR}

Agile RF signal generation and treatment are necessary to allow the creation and use of multi-spectral-hole patterns. An SDR platform is particularly suited for this application as it provides such capability both for emission (TX) and reception (RX) channels, at a moderate cost and with relative ease of programming and reconfiguration, where only knowledge of Python and, in some rare cases, C/C++ programming languages is required.  It is important to note that some optimizations, such as the correct synchronization between the RX and TX channels and latency minimization, are needed for the application of an SDR platform to our research; significant instrumentation notes on the subject are added in Appendix \ref{app:synchRxTx}.

For our SDR platform, we use commercially available open hardware (Ettus Research X310) with open source software development tools (GNU Radio). The heart of this hardware lies in a pre-programmed Field Programmable Gate Array (FPGA), interfaced, on one side, by an ethernet port to a computer, and, to the other side, to analog-to-digital and digital-to-analog converters (ADC and DAC), operating at 200 megasamples per second (MSPS). These ADC and DAC are themselves connected to analog frontends, called ``daugtherboards'', which can be interchanged, typically for use with signals of various carrier frequencies. We use in our applications exclusively the BasicRX and BasicTX daughterboards, which simply provide near direct access to the ADC and DAC through a balun transformer. We note that many other, more complex, daughterboards exist, allowing in particular upconversion to the GHz range by analog in-phase/quadrature (I/Q) mixing with an on-chip programmable sinewave synthesizer, however, in the context of our experiments, compactness is less relevant than direct access to the signal throughout the entire detection chain.

An ensemble of GNU Radio digital processing blocks, connected together through a data flowgraph and controlled by a graphical user interface (GUI), allows the realization of all the signal generation (through the SDR transmission channels, TX) and recording (through the SDR reception channels, RX) as well as the necessary signal processing for our experimental protocols.

\subsection{Generation of Multiple Optical Frequency Modes:\\Multi-channel TX in SDR}

In TX, the SDR receives a stream of data generated by a GNU Radio-made data flow algorithm from the control computer.  This stream rate determines the bandwidth of the desired spectral pattern to be generated (in our use case, typically 20\,MSPS).  This data stream is sent to the FPGA which makes a Digital Up Conversion (DUC), to a programmable carrier frequency, $f_\mathrm{DUC}$. In essence, the FPGA is receiving a stream of samples from the computer, processes it through an interpolating filter to reach 200\,MSPS, and mixes the result at the modulation frequency, generated by a numerically controlled oscillator (NCO). Note the modulation/demodulation frequency can be set to zero, should no actual up/down conversion be required.  In principle, in this implementation the SDR allows us to generate an arbitrary waveform pattern centered around a programmable frequency in the 0-100MHz range, with a bandwidth equal to the Nyquist frequency of the GNU Radio program sample rate.

\subsubsection{TX data processing blocks}
\label{sec:TX}

We use two distinct methods, depending on the intended experimental goal, for generating the control RF signals of the experiment.

The first method is, in essence, time-based, and produces a single tone signal whose frequency can be swept in a controlled way. This function is not available in the standard libraries of GNU Radio; as such, we wrote a custom data processing block ``sweep-frequency'' in C++. Ample documentation exists about writing GNU Radio Blocks, a process which is almost entirely automatized thanks to the gr-modtool utility which is part of the GNU Radio standard implementation. The sweep-frequency block is, at its core, a numerically controlled oscillator which generates a constant amplitude complex stream of samples, for which the phase of each sample is equal to that of the previous one plus an increment, and this increment is itself incremented at each sample by a value proportional to the sweep rate of the frequency expressed in Hz/s. Using such a block, the platform can easily generate frequency ramps of controlled amplitude, sweep rate, and initial frequency and phase.

The second method is frequency based. It produces an ensemble of single tone modes of controlled amplitudes and phases. It is implemented by using the GNU Radio built-in inverse Fast Fourier Transform (iFFT) block, which, when fed with the spectral pattern to realize, generates the corresponding complex time-domain signal data stream. Our typical use of such an approach is for realizing the illumination of the rare-earth-doped crystal with multiple optical frequency tones in parallel. In such applications, the waveform to generate is most easily described, in frequency domain, by a discrete Fourier series composed of zero everywhere except at the places where a mode is to be generated.

It should be noticed that these two approaches are complementary, as one is based in the time domain and the other in the frequency domain. We combine the two signals by numerically multiplying the time samples produced by these two functions resulting in the convolution of the spectra in the frequency domain, whereas adding them simply provides the sum of the two spectra.  With the desired spectrum programmed, the resulting RF signal generated by the TX channel of the SDR is sent to drive the AOM in our setup (Fig. \ref{fig:scheme}) to produce the desired optical spectrum; the production of the optical frequency modes contains some caveats, described in Appendix \ref{app:doublepassAOM}.

{\begin{figure}
    \includegraphics[width=0.33\linewidth]{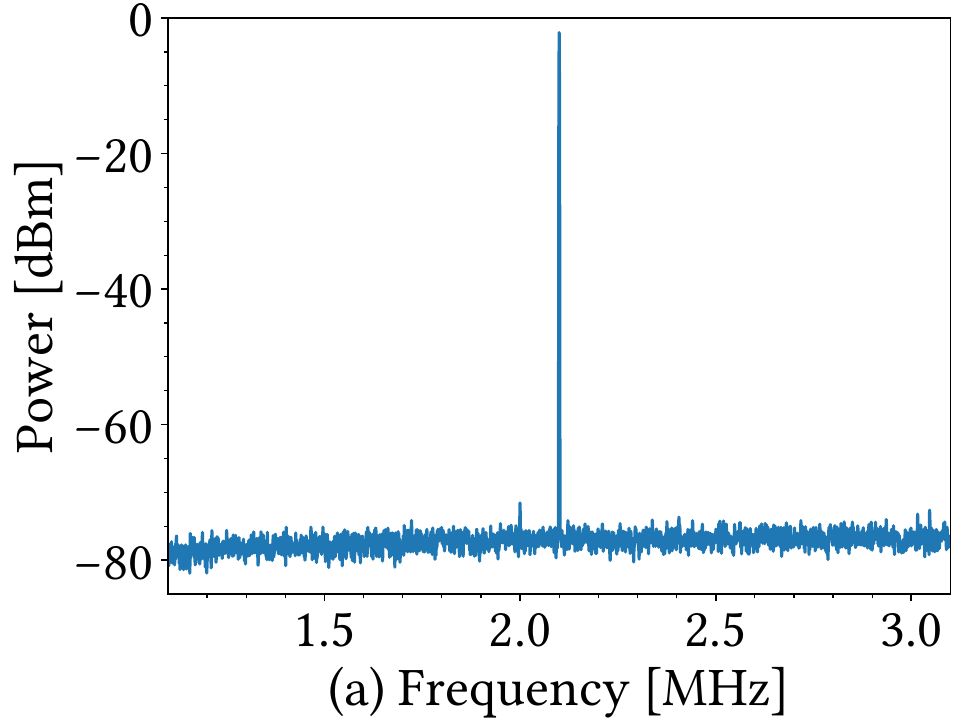}\includegraphics[width=0.33\linewidth]{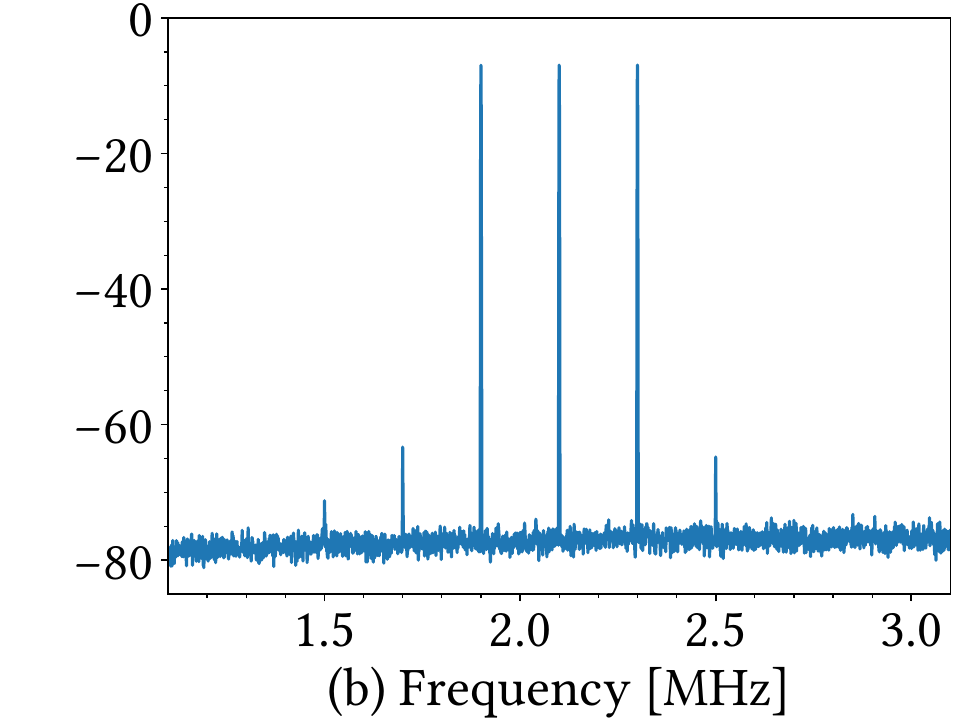}\includegraphics[width=0.33\linewidth]{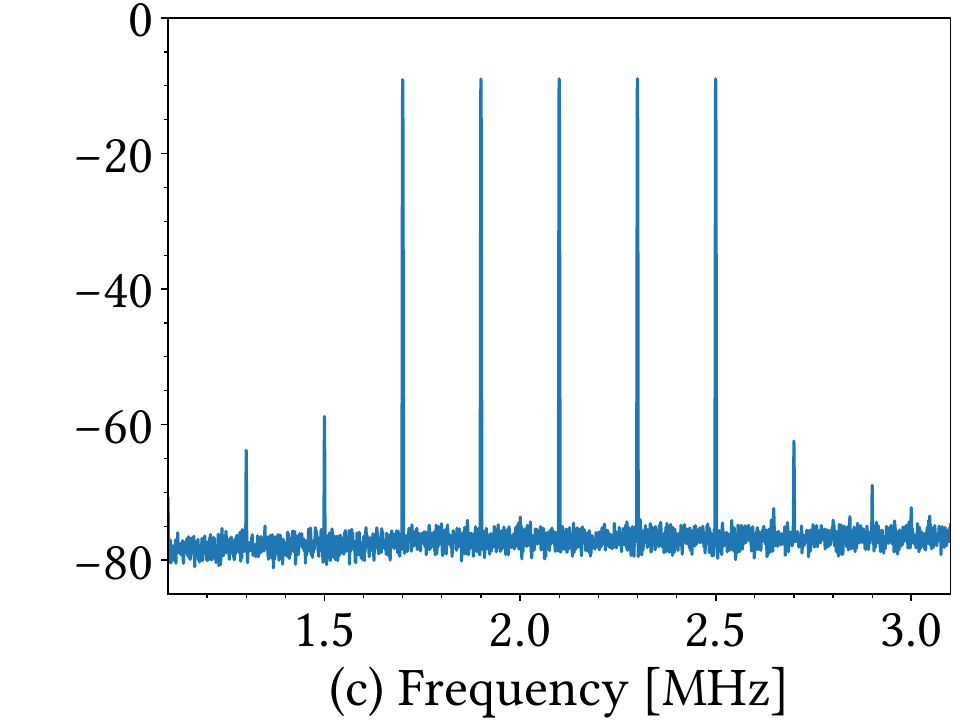}
    \caption{\label{fig:optspectra} Measurements of optical spectra produced in our experiment.  Here, in addition to $1\times$LO-Mode, the RF signal driving the AOM produces $1\times$Probe-mode (a), $3\times$Probe-modes (b), and $5\times$Probe-mode (c). The spectra observed here are the result of beatnotes between the optical LO-mode and the various optical probe modes.}
\end{figure}}

Using these techniques, we demonstrate multi-mode optical spectra generated by our experiment in Fig. \ref{fig:optspectra}.  In this measurement, we simultaneously generate a single fixed-frequency local-oscillator (LO) mode located $f_\mathrm{LO}=2.1\,\mathrm{MHz}$ from the center of the IFFT generated multi-mode spectrum to beat the multi-mode spectrum down to a signal detectable by our APDs. Within each spectrum, the RF signals produced by the individual desired beat notes vary by less than $0.15\,\mathrm{dBm}$ from minimum to maximum translating to a variation in beat-note optical power of less than $2\,\%$.  The origin of the additional sidebands is unknown and effects such as AOM response linearity and RF-chain/detection saturation have been investigated.  Noting the log scale, the sidebands are negligibly small, and, without regard to their magnitude, do not contribute to the error signal, which is generated by demodulation at the specific desired frequencies followed by aggressive filtering.

\subsection{Reception of Multiple Optical Frequency Modes:\\Multi-channel RX in SDR}

In the SDR, the input signal is digitized and then digitally down-converted (DDC) internally by a digitally synthesized demodulation signal of programmable frequency in the 0-100MHz range, $f_\mathrm{DDC}$, referenced to an external 10\,MHz clock.  We set the digital down-conversion frequency to frequency separation between the center of the IFFT multi-mode spectrum and the local oscillator mode so that $f_\mathrm{DDC}=f_\mathrm{LO}$. The resulting baseband signal is streamed to the connected control computer, where a driver (called Universal Hardware Driver ``UHD'') receives it and sends it to a GUI-designed/python-programmed data flow treatment algorithm realized in the GNU Radio framework. We typically use a 2\,MSPS data rate from the FPGA to the computer, which is sufficient for our applications, but have also successfully worked at higher sampling rates. 

\subsubsection{RX data processing blocks}

After DDC, the decomposition of the signals generated by the two photodiodes into their separate frequency channels is realized on the control computer via poly-phase channelizer blocks, a standard component of the GNURadio library \cite{GNURadio_PPC}. Typically, our current implementation uses the input stream of data at 2\,MSPS from a photodiode to generate 10 streams of data at 200\,kSPS per channel, equidistant in frequency-space, separated typically by $f_\mathrm{Sp}=200\,\mathrm{kHz}$. These streams of data correspond to individual probing channels, and will allow simultaneous probing of several narrow or large spectral holes previously photo-imprinted on the crystal. The poly-phase channelizer structure allows efficient algorithmic implementation with a large control over the filtering of channels so as to minimize cross-talk and distortion. As the base of our polyphase channelizer, we typically use a finite impulse response equiripple low-pass filter \cite{GNURadio_Filter_Design} with pass band of 10\,kHz (less than 1\,dB ripple in-band), stopband at 40\,kHz and 120\,dB attenuation out-of-band, which offers sufficient performance for our applications and is implemented in less than 250 taps. Other filters with different performances may of course be used depending on the application. 

For the comparison of the sample heterodyne phase to the reference heterodyne phase, the complex data stream of each individual frequency channel (10 in our current setup) coming from the channelized DDC Xtal signal is divided by the corresponding frequency channel from the similarly channelized DDC Ref signal and the result is separated into amplitude and phase components. In our typical application to multi-spectral-hole spectroscopy, for a frequency channel where light was effectively applied (i.e., from the appropriate TX channel signal), these quantities correspond to the absorption and phase retardation corresponding to the Kramers-Kronig relations resulting from transmission through a narrow spectral hole. Frequency channels where no light was actually applied exhibit only noise-related content and should be discarded.

\section{Multi-channel Low-power Heterodyne Detection}
\label{sec:Multi-ch_lp_hetdet}

In the previous sections we discussed the production (using an SDR-driven AOM) and detection (SDR-digitized/channelized signals from two APDs) of multiple optical frequency modes.  Here we apply this to an interferometric detection scenario where we demonstrate the noise reduction capabilities of this multi-mode interferometry.

\subsection{Avalanche-enhanced Shot Noise}
\begin{figure}[t]
    \centering
    \includegraphics[width=0.9\linewidth]{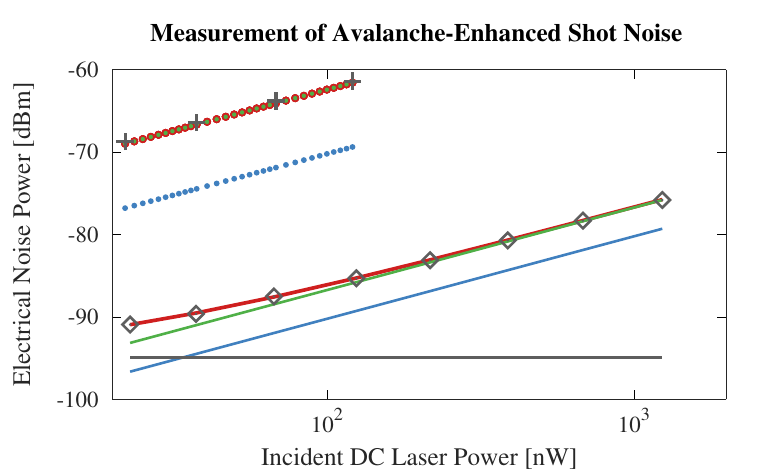}
    \caption{\label{fig:Avalanche_ShotNoise} Measured noise power at different incident laser powers for avalanche multiplication gain $G_\mathrm{A}=10$ (`$\Diamond$' Markers) and for $G_\mathrm{A}=100$ (`$+$' Markers).  The calculated noise contributions for $G_\mathrm{A}=10$ (Solid Lines) and $G_\mathrm{A}=100$ (Dotted Lines) are also plotted.  The photodetector (NEP) noise contribution is plotted as the Black Line. The levels of photon shot noise are plotted in Blue Lines and as avalanche-enhanced shot noise in Green Lines.  The calculated sum total noises are plotted in Red Lines.} 
\end{figure}
In the experiment, we have selected avalanche photodetectors for fast detection of low optical powers.  For this demonstration, we used the Thorlabs APD410A amplified photoreceiver, which has an adjustable avalanche current gain (from $G_\mathrm{A}=10 \rightarrow 100$) and uses a Hamamatsu S12023-10 avalanche photodiode.  This photoreceiver has an electronic noise at the output that varies greatly with measurement frequency.  Around our measurement beat frequency (typically 1\,MHz), the manufacturer specifications indicate an output noise spectral density of $2\times10^{-14}\,\mathrm{V^{2}Hz^{-1}}$ or $-95\,\mathrm{dBm}$ into $50\,\mathrm{\Omega}$ with a $1\,\mathrm{kHz}$ measurement bandwidth. 

For low-power precision interferometry, shot noise is a primary concern. Additionally, avalanche photodetectors manifest a higher level of measured shot noise due to the additional statistical process of the avalanche effect \cite{McIntyre1966:1474241}.  This so-called excess noise factor,
\begin{equation}
     F_\mathrm{x} = {\kappa}{G_\mathrm{A}} + (1-{\kappa})\left(2-\frac{1}{G_\mathrm{A}}\right),
\end{equation}
depends on the ratio of the hole to electron ionization rates, $\kappa$. For the Hamamatsu S12023-10 in the avalanche diode, this value is $\kappa=0.4$ at our measurement wavelength. For an average photon-incidence rate $\bar{n}$ on a detector with quantum efficiency $\eta$, the single sided power spectral density of avalanche-enhanced current shot noise (ASN) is given by
\begin{equation}
     I_{ASN} = 2q_\mathrm{e}^2F_\mathrm{x}G_\mathrm{A}^2\eta\bar{n},
\end{equation} 
where $q_\mathrm{e}$ is the electron charge.  These modeled noise contributions, alongside their sums, are plotted in curves as a function of DC incident power in Fig. \ref{fig:Avalanche_ShotNoise}.  These are in close agreement with the measured noise power plotted with `$\Diamond$' markers (avalanche multiplication set to $G_\mathrm{A}=10$) and with `$+$' markers ($G_\mathrm{A}=100$).  As seen in the figure, when going from a $10\times$ to $100\times$ avalanche gain, the $20\,\mathrm{dB}$ increase in signal is accompanied by a $\geq 23\,\mathrm{dB}$ increase in noise across the power measurement range implying the best SNR case is achieved using the detector adjusted to $G_\mathrm{A}=10$ avalanche gain.

\subsection{Illustrative Case: Single Channel Low-power Heterodyne Detection}
\label{sec:single_probe_case}

\begin{figure}[t]
    \centering
    \hspace{-0.01\textwidth}
    \begin{subfigure}[h]{0.45\textwidth}
        \includegraphics[width=0.98\linewidth]{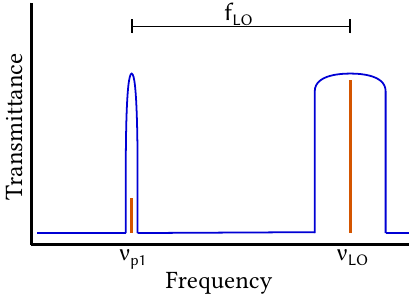}%
        \caption{\label{fig:probe_scheme_1prb0mon}}
    \end{subfigure}   
    \hspace{0.06\textwidth}
    \begin{subfigure}[h]{0.45\textwidth}
        \includegraphics[width=0.98\linewidth]{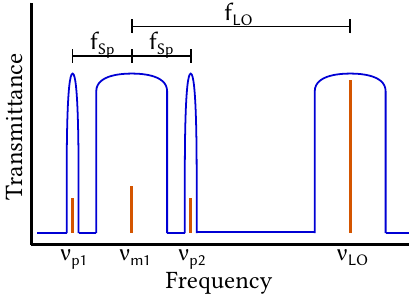}%
        \caption{\label{fig:probe_scheme_2prb1mon}}
    \end{subfigure}
    
    \caption{\label{fig:probe_scheme} Example spectral hole configurations, illustrated in Blue, alongside the measurement laser modes, drawn in Yellow, at the corresponding optical frequencies ($\nu$). Figure \ref{fig:probe_scheme_1prb0mon} consists of one spectral probe mode ($\mathrm{p}_1$) alongside a local-oscillator mode ($\mathrm{LO}$) to form the heterodyne detection frequency, $f_\mathrm{LO}$, as described in Section \ref{sec:single_probe_case}. Figure \ref{fig:probe_scheme_2prb1mon} consists of two spectral probe modes ($\mathrm{p}_1$, $\mathrm{p}_2$) plus one phase noise monitor mode ($\mathrm{m}_1$) alongside the local-oscillator ($\mathrm{LO}$) as described in Section \ref{sec:Complete_Case}.  Note the monitor and local oscillator modes transmit through broad transmission bands to avoid spectroscopic interaction with the sample.}
\end{figure}

In application to the probing of delicate spectral structures (e.g. burned spectral holes) these features necessitate the use of low optical-power spectroscopy.  In a simple illustrative case, we generate a single probe mode ($\mathrm{p}_1$), at optical frequency $\nu_{\mathrm{p}_1}$, such that an optical power $P_{\mathrm{p}_1}=10\,\mathrm{nW}$ is incident on each of the two photodetectors, $\mathrm{PD}^{\mathrm{Xtal}}$ and $\mathrm{PD}^{\mathrm{Ref}}$.   This low-power mode would probe a narrow spectral hole and accumulate a phase proportional to its frequency difference with the spectral-hole center frequency.  Alongside the probe mode, we generate the local-oscillator mode (LO) with power $P_\mathrm{LO}=1\,\mathrm{\mu{W}}$ at optical frequency $\nu_{\mathrm{LO}}$, spaced a distance $f_{\mathrm{LO}}=\nu_{\mathrm{LO}}-\nu_{\mathrm{p}_1}$ from the probe mode, to produce a beat-note at $f_{\mathrm{LO}}$, within the bandwidth of our photodetectors. The LO passes through a broad-band transmissive hole making its phase sensitivity to frequency shifts orders of magnitude lower than the probe narrow-hole signal.  An illustration of this scenario can be seen in Fig. \ref{fig:probe_scheme_1prb0mon}.  In this configuration (as well as all other configurations described later on), the DDC chains of the detection path are set to demodulate against the $f_{\mathrm{LO}}$ frequency, therefore putting the beatnotes between the LO mode and the probe mode at baseband in the GNU Radio data treatment. In our experimental setup, we derive our interferometric signal from the difference between the reference path ${\mathrm{LO}}/{\mathrm{p}_{1}}$ heterodyne beat note phase ($\phi^{\mathrm{Ref}}_{\mathrm{p}_{1}}$) and the crystal path ${\mathrm{LO}}/{\mathrm{p}_{1}}$ beat note phase ($\phi^{\mathrm{Xtal}}_{\mathrm{p}_{1}}$),

\begin{equation}
    \phi_{\mathrm{p}_{1}}(t)=\phi^{\mathrm{Xtal}}_{\mathrm{p}_{1}}(t) - \phi^{\mathrm{Ref}}_{\mathrm{p}_{1}}(t).
\end{equation} 

In this simple configuration, we have seen that technical noise sources begin to exceed the shot noise sensing level at long time scales ($>10\,\mathrm{s}$).   A more complicated probing scheme, which is described in section \ref{sec:Complete_Case}, was demonstrated to effectively subtract these noise sources.

We test the conversion of amplitude to phase noise by intentionally applying, in software, a fixed amplitude modulation of index $m=0.1$ for both the LO and the probe mode in the TX chain, and measuring the resulting phase modulation in the RX detection chain.  In this way, we measured the coefficient of amplitude to phase fluctuations to be $7.1 \times 10^{-4}$\,rad per 100\% amplitude modulation ($m=1$) in our operation regime. This very high level of rejection of amplitude noise, owing largely to the digital approach to our differential measurements and from the naturally small signals on the avalanche photodiodes (thus operating far from saturation) makes this detection scheme largely immune to laser power fluctuations.  

\subsection{Novel Case: Multi-probe-mode Averaging for Reduction of Incoherent Noise Sources}

In the cases where optical beatnote power is restricted by the system being probed (e.g. low power necessary in the probing of delicate spectral structures, such as burned spectral holes), we can improve the phase detection sensitivity by implementing a multi-mode heterodyne detection scheme.  In this case, along side one relatively high power local oscillator (which is removed in frequency space from the spectral features we wish to measure), we create multiple low-power probe modes, separated in frequency space by $f_\mathrm{Sp}$, chosen to occur at the center frequencies of several frequency-separated spectral features.

Over $N$ probe modes, for each individual probe mode, $i$, the phase difference between the reference and sample paths, $\phi_{\mathrm{p}_i}$, is calculated individually and then averaged evenly to find the mean differential phase between the reference and sample paths,

\begin{equation}
    \widebar{\phi_{\mathrm{p}}}(t) = \frac{1}{N} \sum_{i=1}^N \phi_{\mathrm{p}_i}(t). \label{eqn:MeanPhase_PrbOnly}
\end{equation}

The purpose of this exercise is to decrease the detection noise without increasing the power of the individual probe modes.  For some uncorrelated noise power, $S_{\phi_{\mathrm{p}_i}}$, in individual differential phase signals, $\phi_{\mathrm{p}_i}$, the total noise power in the mean differential phase is the sum

\begin{equation}
    S_{\widebar{\phi_{\mathrm{p}}}} = \sum_{i=1}^N \left(\frac{\partial{\widebar{{\phi_{\mathrm{p}}}}}}{\partial{\phi_{\mathrm{p}_i}}}\right)^{2} S_{\phi_{\mathrm{p}_i}}=\frac{1}{N^{2}} \sum_{i=1}^N S_{\phi_{\mathrm{p}_i}}.
\end{equation}

For the typical case where $S_{\phi_{\mathrm{p}_1}}\!\!\!=\!S_{\phi_{\mathrm{p}_2}}\!\!\!=\!\ldots\!=\!S_{\phi_{\mathrm{p}_N}}\!\!\!\equiv\!S_{\phi_{\mathrm{p}}}$  (e.g. shot noise in $N$ equal optical-power probe modes) the expression becomes:

\begin{equation}
    S_{\widebar{{\phi_{\mathrm{p}}}}} =\frac{S_{\phi_{\mathrm{p}}}}{N}.
\end{equation}

\begin{figure}[t]
    \centering
    \includegraphics[width=0.9\linewidth]{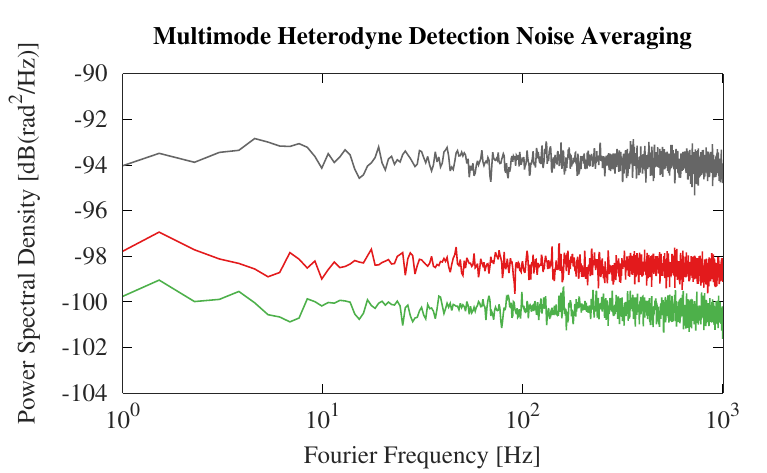}
    \caption{\label{fig:PSD_probe_mode_averaging} The PSD of the detection phase noise.  Averaging the digitally demodulated signals produces the predicted $-4.8\,\mathrm{dB}$ (for $N=3$, red) and $-7.0\,\mathrm{dB}$ (for $N=5$, green) reduction in shot-noise compared to the single mode case (black).}
\end{figure}

We demonstrate this noise reduction by calculating the power spectral densities of the recorded time series of detection noise, that is, differential phase noise between the Ref and Xtal paths.  This phase sensitivity measurement was made in the condition where no spectroscopic sample is in place (i.e. Ref path is through freespace and the Xtal path is through the cryostat, but with no crystal installed).  For the case of a single probe mode ($N=1$), we measured a mean power spectral density of detection noise dominated by avalanche-enhanced shot noise at $S_{\phi_{\mathrm{p}}} = -93.9\,\mathrm{dB({rad^{2}}{Hz^{-1}})}$ in the Fourier frequency range from $10\,\mathrm{Hz}$ to $1\,\mathrm{kHz}$.  The cases of $N=3$ and $N=5$ probe modes are expected to result in a $10\log_{10}(\frac{1}{3})\approx-4.8\,\mathrm{dB}$ and $10\log_{10}(\frac{1}{5})\approx-7.0\,\mathrm{dB}$ reduction in detection noise compared to the single probe-mode case, respectively.  These are confirmed by our measurements, plotted in Fig. \ref{fig:PSD_probe_mode_averaging}.

\subsection{Complete Case: Reducing Differential Path Noise Using a Monitor Mode}
\label{sec:Complete_Case}

The real power of our platform is revealed when utilizing its multiple detection channels in a more complex way. Of particular interest is the use of dedicated channels for the suppression of common-mode differential path noise (CMDN), that is, phase noise which is differential between the two separate optical paths but common amongst the modes within each path.  As sources of CMDN are present and correlated in each of the probe modes, they sum coherently as we average over several probe modes, and subsequently do not reduce when averaging over probe modes.  For this we take a different approach, adding additional ``Monitor Modes'' ($\mathrm{m}_i$) monitoring differential path noise.  We produce the monitor modes using the same method as the probe modes, and, like the probe modes, they are digitally demodulated in the SDR after detection.  By adding monitor modes at frequencies distanced from the spectral features of interest, we are able to measure undesired differential phase fluctuations between the reference and sample paths independent from the desired signal, the spectroscopic phase contribution.  An illustration of the generated laser spectrum (yellow) is laid against an illustration of a spectral hole pattern (blue) in Fig. \ref{fig:probe_scheme_2prb1mon}.  In this example, two probe modes are generated at the location of two narrow line spectral holes, while a single monitor mode is produced between them at the location of a broad transmissive hole.  Narrow holes are pre-burned by selectively pumping the ions at a particular resonance frequency using a relatively high-power narrow-linewidth laser, frequency stabilized to a cavity, over a short time (of order $1\,\mathrm{{\mu}{W}}$ over $1\,\mathrm{{s}}$); a broad spectral hole is produced by sweeping the hole-burning laser over the desired frequency range to burn.  A local oscillator, which brings these from optical frequencies to a measurable RF frequency, transmits through a broad spectral hole at a distance $f_{\mathrm{LO}}=\nu_\mathrm{LO}-\nu_{\mathrm{m}_1}$ from the monitor mode.

Sources of CMDN include differential optical path length noise and phase noise resulting from unwanted on-axis interference from undesired reflections at surfaces at normal incidence.  In our experiment, the sample path contains several opportunities for these unwanted Fabry-Perot effects as the sample is housed within several thermal shields inside a vacuum chamber, each with a glass window to transmit the sample beam.  These parasitic etalon effects are of particular concern as the resulting phase noise has a frequency dependence, causing an imperfect cancellation when demodulating against the local-oscillator mode alone, which is limited to a cancellation factor set by the ratio of the beat frequency to optical frequency.  This prompts us to use a monitor mode centered in frequency space between probe modes. When using an even number of probe modes, this symmetric configuration in frequency space of the probe/monitor modes allows for a more complete first-order cancellation of the phase noise resulting from this parasitic interference, a detail discussed further in Appendix \ref{app:parsiticFP}.  Furthermore, a common monitor mode measures and allows us to subtract additional differential phase noise between the channels, such as timing differences between the two ADC channels and electrical signal delay between the photodetectors.

\begin{figure}[t]
    \centering
    \includegraphics[width=0.98\linewidth]{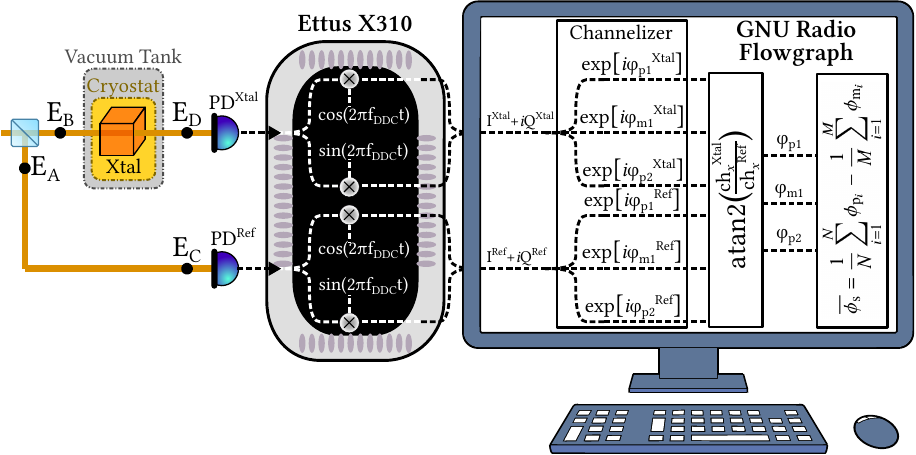}%
    \caption{\label{fig:signal_flow} A simplified diagram of the RX signal detection in the experiment for the case of 4 modes: $2\times$probe-mode, $1\times$monitor-mode, $1\times$LO-mode case illustrated in Fig. \ref{fig:probe_scheme_2prb1mon}. The details are described in the text.}
\end{figure}

An illustration of the signal flow for this case (Fig. \ref{fig:signal_flow}) can demonstrate the noise subtraction using a monitor mode.  Following this signal flow, the laser field in the reference path accumulates phase according to its optical path before being detected at $\mathrm{PD}^{\mathrm{Ref}}$, while the laser field in the sample path additionally accumulates the spectroscopic signal phase and is detected at $\mathrm{PD}^{\mathrm{Xtal}}$.  These signals are digitized separately and digitally demodulated at  $f_\mathrm{DDC}=f_\mathrm{LO}$, bringing the signals into baseband around the center frequency signal, i.e. the monitor mode.  The I-Q signals are passed to the control program on the computer where `Xtal' path signals and `Ref' path signals are filtered by the Polyphase Channelizer into their frequency components.  The phase difference, $\phi_{x} = \phi^{\mathrm{Xtal}}_{x}-\phi^{\mathrm{Ref}}_{x}$, is calculated for each corresponding component ($x=\mathrm{p}_1,\mathrm{m}_1,\mathrm{p}_2$). Using the channelized monitor phase signals, the mean over $M$ accumulated differential monitor-mode phases ($\phi_{\mathrm{m}_i}$) can be subtracted from the simple differential probe-mode phase average (previous case given in Eqn. \eqref{eqn:MeanPhase_PrbOnly}) in realtime, leaving only the average of the desired signal phases:

\begin{equation}
  \widebar{\phi_{\mathrm{s}}}(t) = \frac{1}{N} \sum_{i=1}^N \phi_{\mathrm{p}_i}(t) - \frac{1}{M} \sum_{i=1}^M \phi_{\mathrm{m}_i}(t). \label{eqn:MeanPhase_PrbMon}
\end{equation}

%\widebar{{\phi_\mathrm{s}}}= \frac{1}{N} \sum_{i=1}^N \phi_{\mathrm{p}_i} - \frac{1}{M} \sum_{i=1}^M \phi_{\mathrm{m}_i} \label{eqn:MeanPhase_PrbMon}

\begin{figure}[t]
    \centering
    \includegraphics[width=0.9\linewidth]{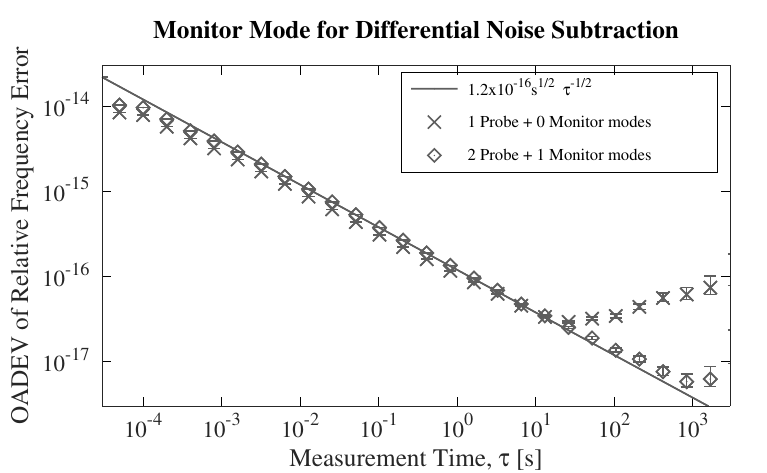}%
    \caption{\label{fig:adev_psd_3h_1MHz} Overlapping Allan deviation measurement of phase fluctuations projected as frequency error w.r.t. average center frequency of narrow spectral holes with typical discriminant $D_{\phi/\nu}=300\,\mathrm{{\mu}{rad}}\,\mathrm{Hz}^{-1}$. Compared to the single mode case (`$\times$' Markers), the inclusion of a monitor mode to subtract differential path length noise reduces sensing noise by an order of magnitude at $10^3\,\mathrm{s}$ of integration time (`$\Diamond$' Markers).  For visual reference, a line $\propto\tau^{-1/2}$ has been included on the plot.}
\end{figure}

We see the result of CMDN best in long phase noise measurements.  To demonstrate the baseline measurement noise, a three hour long Allan deviation of the Ref-Xtal differential phase, which included the vacuum chamber and thermal shields (but no spectroscopy samples) in the sample path, is plotted in Fig. \ref{fig:adev_psd_3h_1MHz}.  Here we projected the measured phase noise as fractional frequency difference between a single probe mode and the center frequency of a hypothetical spectral hole with a typical frequency discriminant $D_{\phi/\nu}=300\,\mathrm{{\mu}{rad}}\,\mathrm{Hz}^{-1}$. A measurement of phase noise using a single probe mode beat down with a local-oscillator mode is plotted with `$\times$' markers (the case described in Section \ref{sec:single_probe_case}).  Here we see detection white noise at high frequencies dominates the short averaging-time ($\tau$) measurement points, rolling off as $\tau^{-1/2}$ until about $10\,\mathrm{seconds}$ where sources of CMDN take over at low frequencies/long integration times.  A measurement that utilized a monitor mode is plotted with `$\Diamond$' markers, this result was obtained using one local oscillator mode, one monitor mode, and two probe modes (the example illustrated in Fig \ref{fig:probe_scheme_2prb1mon}).  Using the monitor mode to subtract CMDN, we obtain an order of magnitude reduction in sensing noise at long timescales.

\subsubsection{Comments on detection noise using a monitor mode}

Note in Eqn. \eqref{eqn:MeanPhase_PrbMon} that, for successful subtraction of CMDN we subtract the monitor mode outside of the averaging of the probe modes, thus it becomes an additional contribution to shot-noise.  The total noise power can be described by the sum of uncorrelated noise powers of the individual modes,
\begin{align}
    S_{\widebar{{\phi_{\mathrm{s}}}}} &= \sum_{i=1}^N \left(\frac{\partial{\widebar{{\phi_{\mathrm{s}}}}}}{\partial{\phi_{\mathrm{p}_i}}}\right)^{2}S_{\phi_{\mathrm{p}_i}} + \sum_{i=1}^M \left(\frac{\partial{\widebar{{\phi_{\mathrm{s}}}}}}{\partial{\phi_{\mathrm{m}_i}}}\right)^{2}S_{\phi_{\mathrm{m}_i}}\\&=\frac{1}{N^{2}} \sum_{i=1}^N S_{\phi_{\mathrm{p}_i}}+\frac{1}{M^{2}} \sum_{i=1}^M S_{\phi_{\mathrm{m}_i}}.
\end{align}
For the case where probe modes have equal noise power ($S_{\phi_{\mathrm{p}_1}}\!\!\!=\!S_{\phi_{\mathrm{p}_2}}\!\!\!=\!\ldots\!=\!S_{\phi_{\mathrm{p}_N}}\!\!\!\equiv\!S_{\phi_{\mathrm{p}}}$) and monitor modes have equal noise power ($S_{\phi_{\mathrm{m}_1}}\!\!\!=\!S_{\phi_{\mathrm{m}_2}}\!\!\!=\!\ldots\!=\!S_{\phi_{\mathrm{m}_M}}\!\!\!\equiv\!S_{\phi_{\mathrm{m}}}$) the expression becomes:
\begin{equation}
    S_{\widebar{{\phi_{\mathrm{s}}}}} =\frac{S_{\phi_{\mathrm{p}}}}{N} + \frac{S_{\phi_{\mathrm{m}}}}{M}.
    \label{eqn:ProbMonNoisePSD}
\end{equation}

For the case of shot-noise limited detection, the power spectral density of phase noise in each individual probe or monitor mode,
\begin{equation}
    S_{\phi} = F_\mathrm{x}\frac{{h}{\nu}}{4{\eta}{\bar{P}}}, 
\end{equation}
is inversely proportional to the mean power, $\bar{P}$, in that mode.  If we have individual probe modes of equal power ($\widebar{P_{\mathrm{p}_1}}\!\!\!=\!\widebar{P_{\mathrm{p}_2}}\!\!\!=\!\ldots\!=\!\widebar{P_{\mathrm{p}_N}}\!\!\!\equiv\!\widebar{P_{\mathrm{p}}}$) and monitor modes of equal power ($\widebar{P_{\mathrm{m}_1}}\!\!\!=\!\widebar{P_{\mathrm{m}_2}}\!\!\!=\!\ldots\!=\!\widebar{P_{\mathrm{m}_M}}\!\!\!\equiv\!\widebar{P_{\mathrm{m}}}$) the total noise spectral density is then given by
\begin{equation}
    S_{\widebar{{\phi_{\mathrm{s}}}}} =F_\mathrm{x}\frac{{h}{\nu}}{4{\eta}}\left(\frac{1}{{N}{\widebar{P_{\mathrm{p}}}}} + \frac{1}{{M}{\widebar{P_{\mathrm{m}}}}}\right).
    \label{eqn:ShotNoisePSD}
\end{equation}
The shot noise contribution from monitor mode can be made arbitrarily small by fixing ${M}{\widebar{P_{\mathrm{m}}}}\gg{N}{\widebar{P_{\mathrm{p}}}}$.  However, if we have a total working power limit, $P_\mathrm{t}$, set by either the available laser power for mode production or by the saturation limit of the detectors, to be split amongst all the probe/monitor modes such that $P_\mathrm{t} = {N}{\widebar{P_{\mathrm{p}}}} + {M}{\widebar{P_{\mathrm{m}}}}$, we see that Eqn. \eqref{eqn:ShotNoisePSD} is minimized when ${N}{\widebar{P_{\mathrm{p}}}} = {M}{\widebar{P_{\mathrm{m}}}}$.  Additionally, in the application to SHB, we wish to limit the power in the probe and monitor modes to avoid `overburning', or the spectral distortion of the spectral holes.

Returning to the example of probing burned spectral holes, it is impractical to make several broad transmissive spectral holes to allow multiple monitor modes, and, in fact, it is mostly the probe-mode power that is limited by the delicate narrow spectral features, much more so than a monitor mode which interacts with a broad spectral feature. Thus it is possible to fix the number of monitor modes to $M=1$ and minimize Eqn. \eqref{eqn:ShotNoisePSD} with ${\widebar{P_{\mathrm{m}}}} = N{\widebar{P_{\mathrm{p}}}}$.  

\begin{figure}[t]
    \centering
    \vspace{-2ex}
    \includegraphics[width=0.9\linewidth]{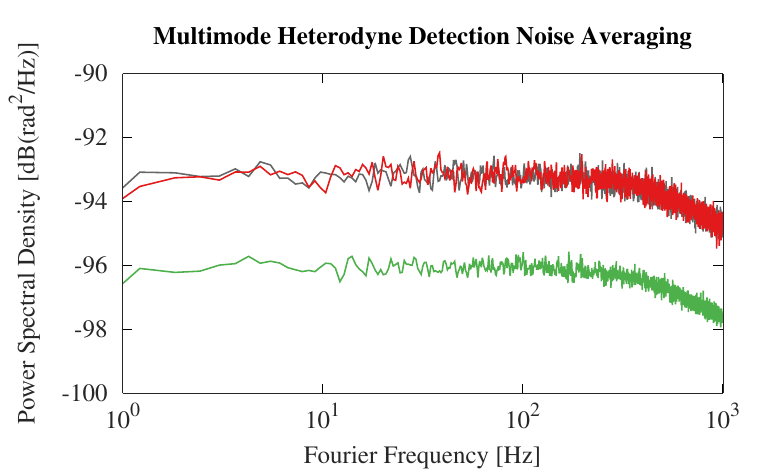}\vspace{-1ex}
    \caption{\label{fig:PSD_PrbMon_mode_averaging} The PSD of the detection phase noise using a monitor mode for CMDN subtraction.  The Black curve represents Case I, the Red curve represents Case II and the Green curve represents Case III, described in the text. \vspace{-3ex}}
\end{figure}

We demonstrate these results in measurements of the noise power spectral density for three cases:\\
Case I: One single probe mode (N=1) of power $\widebar{P_{\mathrm{p}}} = 10\,\mathrm{nW}$ and no monitor mode (black)\\
Case II: Two probe modes (N=2) and one monitor mode ($\widebar{P_{\mathrm{m}}} = 2\widebar{P_{\mathrm{p}}}= 20\,\mathrm{nW}$) (red)\\
Case III: Four probe modes (N=4) and one monitor mode ($\widebar{P_{\mathrm{m}}} = 4\widebar{P_{\mathrm{p}}}= 40\,\mathrm{nW}$) (green)\\
It is clear from Eqn. \eqref{eqn:ProbMonNoisePSD} and Eqn. \eqref{eqn:ShotNoisePSD} that moving from Case I to Case II the shot-noise contribution from the probe modes are halved, however we add an equal contribution from a monitor mode resulting in the same shot-noise level for both cases.  Moving to Case III we would then expect a $10\log_{10}(\frac{2}{4})\approx-3\,\mathrm{dB}$ reduction in shot-noise.  These are confirmed by our measurements plotted in Fig. \ref{fig:PSD_PrbMon_mode_averaging}.  For the long-term Allan deviation measurement (Fig. \ref{fig:adev_psd_3h_1MHz}) the cases correspond roughly to Case I (1 probe + 0 monitor) and Case II (2 probe + 1 monitor), however, with the difference being for that measurement we did not increase the power in the monitor mode (i.e. $\widebar{P_{\mathrm{m}}} = \widebar{P_{\mathrm{p}}}= 10\,\mathrm{nW}$) resulting in a slightly higher level of shot noise ($\propto \tau^{-1/2}$) in the 2 probe + 1 monitor mode measurement in Fig. \ref{fig:adev_psd_3h_1MHz}.

\section{Conclusion}

In the context of our SHB laser stabilization experiment, we have demonstrated a scheme for flexible multi-mode heterodyne interferometry.  The scheme's agility and ease of implementation derive from the use of open-source hardware (Ettus USRP SDR) and software (GNU Radio) platforms.  We have shown an adjustable interrogating setup and detection noise reduction that allows us to work around the power-limit restrictions of spectral hole probing.  We have analysed the detection noise performances in the system and the results suggest that the simultaneous probing ability and the noise performance of this novel technique show the potential to meet the requirement of the SHB-based ultra-stable lasers. Additionally, the strategic utilization of additional monitor modes enables the subtraction of technical noise sources, such as parasitic interferometric effects. We are confident that these techniques and platform will bring benefits not only to our ultra-stable laser experiment but also to any spectroscopy or heterodyne interferometry schemes that require agile frequency control or encounter power-per-frequency mode limitations (e.g., laser locking to iodine\cite{philippe_frequency_2016}). The relative ease of use of the SDR-based platform has allowed a full noise characterization of the system and tests of spectral hole locking with a unity gain frequency $\approx100\,\mathrm{Hz}$.  Future work will involve investigating ways to decrease signal delay in the system to improve locking bandwidth.

\appendix
\section{Appendix: Synchronisation Between RX and TX, SDR-based Servoloops, and Other Optimizations}
\label{app:synchRxTx}

For direct analysis of the transmission heterodyne signal the SDR and GNURadio-based platform is very well suited as, in addition to saving data for post processing, it allows direct access to the multiple channels of information. However, it should be mentioned, that with no special care taken, the time offset and thus relative phase between the two RX channels is random (although constant over a given run of the program). In order to provide a minimal and reproducible phase offset between the two inputs of the SDR platform, it is necessary to synchronize their DDC chains. This is done simply in software at the beginning of the program, by using timed commands to set the two input channels' DDC chain parameters simultaneously. The corresponding Python code is:
\begin{lstlisting}
    now = self.USRP_RX.get_time_now()
    cmd_time = now + uhd.time_spec(1) 
    self.USRP_RX.set_command_time(cmd_time)
    self.USRP_RX.set_center_freq(freq, 0) 
    self.USRP_RX.set_center_freq(freq, 1) 
    self.USRP_RX.clear_command_time()
\end{lstlisting}
where the \verb|USRP_RX| object is instantiated in the flowgraph by a ``UHD: USRP Source'' block and \verb|freq| is the demodulation frequency of the DDC chain of the Ettus X310.

Similarly, in closed-loop implementations where the error signal that is produced by the multichannel heterodyning is used to generate a correction signal to be emitted by one of the TX channels, it is desirable to minimize the time of propagation of the signal between the RX and TX channels. Furthermore, making this propagation time repeatable is also desirable as otherwise the in-loop servo parameters would need to be fine tuned at each run of the program for optimal performance (defined in particular as maximal gain and bandwidth while maintaining servo-loop stability). The GNURadio framework is not particularly well suited for this application as it is largely designed to optimize throughput, not latency. However, it is possible to implement a closed-loop application with acceptable performance, at least for testing and prototyping, provided special care is taken to force the scheduler to accommodate this non-standard requirement.

By default, the GNURadio implementation we use (version 3.8.4.0 compiled from the GitHub sources) sets a 100\,ms delay time between the start of the RX digitization and that of the TX generation. As a consequence, this imposes a maximal locking bandwidth of the order of 2.5\,Hz, no matter how optimized the flowgraph may be. It is therefore necessary to modify this default behavior, which can be achieved easily with the following piece of code run at the time of initialization:
\begin{lstlisting}
    now = self.USRP_RX.get_time_now()
    # prepare RXTX sync 1s from now
    t0 = now + uhd.time_spec(1) 
    dt = uhd.time_spec(delay_time)
    self.USRP_RX.set_start_time(t0) 
    self.USRP_TX.set_start_time(t0+dt) 
\end{lstlisting}
where, as before, the \verb|USRP_RX| and \verb|USRP_TX| objects are instantiated in the flowgraph by the ``UHD: USRP Source'' and ``UHD: USRP Sink'' blocks, respectively.  The \verb|delay_time| parameter should be set to the smallest value possible before triggering failures of the scheduler to meet the timing requirement (such failure is readily observed by repeated occurrence of late packet events).  Note that it is possible to set the \verb|delay_time| parameter to zero, by providing an extra delay block of fixed delay value in the flowgraph. This somewhat surprising behavior is related to specifics of the GNURadio framework and delay block programming, and may not necessarily be expected to be true for all versions. 

Once this is done, it then becomes useful to optimize the working of the flowgraph to allow decreasing the latency as much as available computing power allows. We use a computer equipped with an Intel Xeon W-2275 multi-core processor. We found it beneficial to de-activate hyperthreading and force the operating clock frequency of all cores to the maximum availability (4.0\,GHz in our case, using the Linux cpu-freq-set package).  Additionally, using the isol-cpu Linux package, we isolate 12 out of the 14 cores, making them available only to the GNURadio blocks involved in the closed loop part of the flowgraph. We have also found it beneficial to assign each GNURadio block within the servo-loop signal path to a specific processor core (whereas in normal operation, the Linux scheduler would be free to assign and re-assign dynamically the blocks' threads to the various cores at run-time). 

Utilizing this approach, with careful optimization of core use by each GNURadio block, we are able to achieve a 100\,Hz feedback bandwidth with our complete multichannel flowgraph (10 optical frequency channels), consistent with the minimal value of 2\,ms of the \verb|delay_time| without significant package loss or late arrival. Future work in our experiment involves the implementation of a custom purpose-built FPGA system, in place of the GNU Radio/SDR system presented here, in order to reduce signal latency.  We will soon implement this change in our setup where we expect the latency reduction to translate into a higher locking bandwidth in our spectral-hole-based ultra-stable laser.

\section{Appendix: Managing Multiple Frequency Components in a Double Pass Acousto-optic Modulator}
\label{app:doublepassAOM}

We imprint the frequency modes on the laser beam using an AOM in a double-pass configuration to produce an optical beam whose geometrical properties do not depend on the RF spectral pattern which drives the AOM. However, an RF driving signal with multiple frequency components results in an intermodulation between the modes upon the second pass through the AOM. Fortunately, as described in \cite{Jobez2016:PhysRevA.93.032327}, a simple algorithmic solution exists to counteract these effects. This can be seen mathematically if one writes, for a given monochromatic input laser field ($A_{\rm in} \exp{[{\imi} 2 \pi \nu_0 t]}$) of amplitude $A_{\rm in}$ and frequency $\nu_0$, entering the double-pass AOM driven by an RF field ($A_{\rm RF}(t) \exp{[{\imi} (2\pi f_0 t + \phi_{\rm RF}(t) )]}$) at a carrier frequency $f_0$ and time-varying amplitude $A_{\rm RF}(t)$ and phase $\phi_{\rm RF}(t)$, the resulting field, $A_\mathrm{out}$, at the output of the first-order diffracted double pass AOM is proportional to the square of the driving field, $A_\mathrm{RF}$:
\begin{equation}
    A_\mathrm{out} \propto A_{\rm in} {\eue}^{{\imi} 2 \pi \nu_0 t} \cdot A_{\rm RF}^2(t) {\eue}^{{\imi}[4\pi f_0 t + 2\phi_{\rm RF}(t)]}.
\end{equation}
Note that we have omitted the position dependence of the travelling wave here.

The production of unwanted cross-terms can therefore be avoided by simply taking the complex-square root of each time sample before sending it to the AOM RF driving field, a simple-enough process in an SDR platform. Note, however, that this simple mathematical expression requires the AOM to operate in its \textit{linear regime} (i.e. driven at an RF power significantly lower than that giving maximum diffraction efficiency in CW operation).  This is an acceptable constraint in our application where we require only a very low optical power (of order $\mathrm{\mu W}$) for probing the spectral holes, three orders of magnitude less than our available source laser.  Note, also, that we need to use a complex-number implementation of the square root function. For this, we have chosen to decompose the complex stream into its amplitude and phase; in parallel we divide the phase by 2 and take the real square root of the amplitude, finally recombining the resulting values to obtain the complex square root of the input sample. 

With more complex spectral patterns being desired for experimentation, an extra subtlety appears. The phase data in a digital system is typically wrapped in the range of $-\pi$ to $\pi$. However, in our data treatment, after passing through the complex square root function, the phase value will become half of the original one, which will be in the range $-\pi/2$ to $\pi/2$. In the complex plane, two successive samples $A\exp{[{\imi} (\pi-\epsilon)]}$ and $A\exp{[{\imi} (\pi+\epsilon)]}=A\exp{[{\imi} (-\pi+\epsilon)]}$, which are continuous in phase for very small $\epsilon$, will therefore produce, after complex-square-rooting, two successive samples $A\exp{[{\imi} (\pi/2-\epsilon/2)]}$ and $A\exp{[{\imi} (-\pi/2+\epsilon/2)]}$. Although the squaring effect of the double pass AOM is mathematically canceled out, the physical signal produced will exhibit a sudden jump in phase at this point (except if the amplitude happens to be exactly zero at that moment, which is for example the case for simple spectra which are symmetric with reference to the carrier frequency). Such a sudden jump is not perfectly realizable by the AOM which effectively acts as a low pass filter, leading to distortion of the signal. To overcome this problem, we wrote a ``phase re-wrap'' function to ensure that the phase data wrapped in the $-\pi$ to $\pi$ interval is re-wrapped in the range of $-2\pi$ to $2\pi$ before the calculation of the complex square root.

\section{Appendix: Cancellation at First Order of Parasitic Fabry-Perot Cavity Effects for Appropriate Multi-mode Probing}
\label{app:parsiticFP}

In a spectral hole configuration described in Fig. \ref{fig:probe_scheme_2prb1mon} and section \ref{sec:Complete_Case}, the particular use of narrow and broad spectral holes allows generating an error signal which is largely immune to parasitic Fabry-Perot etalon effects: effectively, the optical beatnote between the LO mode and the monitor mode provides a measurement of the phase retardation due to the parasitic Fabry-Perot effect, while the two symmetric probe modes, subject to a similar parasitic Fabry-Perot etalon effect, are also sensitive to the phase change due to the narrow spectral holes. An appropriate combination of the phases of these different beatnotes provides a signal suitable to frequency lock a laser onto the central frequency of the narrow holes with removal of the parasitic Fabry-Perot etalon effect. The beatnote signal between the LO mode and any of the other three modes exhibits a component related to the dispersion of the spectral hole pattern in addition to a component related to parasitic Fabry-Perot effect. 

Considering, for simplicity but without loss of generality, a single Fabry-Perot parasitic cavity of length $L$ in vacuum, it comes, from the well known complex transmission coefficient $T/(1-Re^{i.4{\pi}{\nu}{L}/c})$ (where $\nu$ is the optical frequency, $c$ the speed of light and $T$ and $R$ are the combined transmissivity and reflectivity of the mirrors - assumed to be independent of $\nu$):
\begin{equation}
    \phi_i-\phi_{\rm LO} = \arctan \left[ \frac{R \sin(4 \pi L \nu_i /c)}{1+R \cos(4 \pi L \nu_i /c)}\right]
        - \arctan \left[ \frac{R \sin(4 \pi L \nu_{\rm LO} /c)}{1+R \cos(4 \pi L \nu_{\rm LO} /c)}\right],
\end{equation} 
where $i=\{ {\rm p_1}, {\rm m_1}, {\rm p_2}\}$ for the case of two probe modes and one monitor mode (ie. the case used in Fig. \ref{fig:adev_psd_3h_1MHz}, the measurement parasitic Fabry-Perot induced noise cancellation). For a parasitic effect, we can consider $R \ll 1$, leading to 
\begin{equation}
    \phi_i \simeq R \sin(4 \pi L \nu_i /c).
\end{equation} 
Therefore, combining the measured phase of the beatnotes between the LO-mode and the $i$-modes which are spectrally separated by $\Delta \nu$, with relative weighing coefficients (1/2, -1, 1/2) which provides an error signal suitable to lock the probing laser on the central frequencies of the narrow spectral holes, we get:
\begin{eqnarray}
    \epsilon_{\rm FP} & = & \frac{1}{2}(\phi_{\rm p_1}-\phi_{\rm LO}) - (\phi_{\rm m_1}-\phi_{\rm LO}) + \frac{1}{2}(\phi_{\rm p_2}-\phi_{\rm LO}) \\
        & \simeq & R \sin(4 \pi L \nu_{\rm m}/c) [\cos (4 \pi L \Delta \nu/c)-1].
\end{eqnarray}

Because $\Delta \nu \ll c/L $ for practical experimental implementations (in a table top experiment the largest $L$ may be of order $1\,\mathrm{m}$ whereas in our SDR-based system, $\Delta \nu$ is typically of the order of a few 100\,kHz), this strongly reduces the effect of the parasitic Fabry-Perot cavity compared to single channel probing (in fact, it cancels out at first order in $\Delta \nu$). Note that using a more complex (non-linear) combination of the phases of probe and monitor modes (with an increased number of monitor modes), it is also possible to cancel it at higher orders as well (effectively using the phase of the monitor modes to fit a polynomial estimation of the parasitic Fabry Perot effect in the vicinity of the probe modes and subtracting this effect from the measured probe-mode phases), should the necessity arise in a particular highly-demanding application.

\begin{backmatter}
\bmsection{Funding}
The project has received financial support from Ville de Paris Emergence Program, the Région Ile de France DIM C’nano and SIRTEQ, the LABEX Cluster of Excellence FIRST-TF (ANR-10-LABX-48-01) within the Program “Investissement d’Avenir” operated by the French National Research Agency (ANR), and the 15SIB03 OC18 and 20FUN08 NEXTLASERS projects from the EMPIR program cofinanced by the Participating States and from the European Union’s Horizon 2020 research and innovation program, and the UltraStabLaserViaSHB (GAP-101068547) from Marie Skłodowska-Curie Actions (HORIZON-TMA-MSCA-PF-EF) from the European Commission Horizon Europe Framework Programme (HORIZON).

\bmsection{Acknowledgments}
We thank D. Nicolodi for the use of his custom GNURadio blocks, on which we based our sweep-frequency block.

\bmsection{Disclosures}
The authors declare no conflicts of interest.

\bmsection{Data availability} Data underlying the results presented in this paper are available in \cite{Multi-mode_Heterodyne_Laser_Interferometry_Realized_via_Software_Defined_Radio:doi:10.5281/zenodo.8390942}

\end{backmatter}

%%%%%%%%%%%%%%%%%%%%%%% References %%%%%%%%%%%%%%%%%%%%%%%%%

%%%%%%%%%% If using BibTeX:

\bibliography{ReferencesSHBPaper}

\begin{thebibliography}{10}
\newcommand{\enquote}[1]{``#1''}

\bibitem{1075396}
T.~Okoshi, \enquote{Recent advances in coherent optical fiber communication
  systems,} {\protect\JournalTitle{Journal of Lightwave Technology}}
  \textbf{5}, 44--52 (1987).

\bibitem{TheLaseranditsApplicationtoMeteorology}
G.~G. Goyer and R.~Watson, \enquote{The laser and its application to
  meteorology,} {\protect\JournalTitle{Bulletin of the American Meteorological
  Society}} \textbf{44}, 564 -- 570 (1963).

\bibitem{McElroy:72}
J.~H. McElroy, \enquote{Infrared heterodyne solar radiometry,}
  {\protect\JournalTitle{Appl. Opt.}} \textbf{11}, 1619--1622 (1972).

\bibitem{Abbas:76}
M.~M. Abbas, M.~J. Mumma, T.~Kostiuk, and D.~Buhl, \enquote{Sensitivity limits
  of an infrared heterodyne spectrometer for astrophysical applications,}
  {\protect\JournalTitle{Appl. Opt.}} \textbf{15}, 427--436 (1976).

\bibitem{PhysRevD.92.022004}
J.~Eichholz, D.~B. Tanner, and G.~Mueller, \enquote{Heterodyne laser frequency
  stabilization for long baseline optical interferometry in space-based
  gravitational wave detectors,} {\protect\JournalTitle{Phys. Rev. D}}
  \textbf{92}, 022004 (2015).

\bibitem{Gobron:17}
O.~Gobron, K.~Jung, N.~Galland, K.~Predehl, R.~L. Targat, A.~Ferrier,
  P.~Goldner, S.~Seidelin, and Y.~{Le Coq}, \enquote{Dispersive heterodyne
  probing method for laser frequency stabilization based on spectral hole
  burning in rare-earth doped crystals,} {\protect\JournalTitle{Opt. Express}}
  \textbf{25}, 15539--15548 (2017).

\bibitem{Galland:20}
N.~Galland, N.~Lu\v{c}i\'{c}, S.~Zhang, H.~Alvarez-Martinez, R.~L. Targat,
  A.~Ferrier, P.~Goldner, B.~Fang, S.~Seidelin, and Y.~{Le Coq},
  \enquote{Double-heterodyne probing for an ultra-stable laser based on
  spectral hole burning in a rare-earth-doped crystal,}
  {\protect\JournalTitle{Opt. Lett.}} \textbf{45}, 1930--1933 (2020).

\bibitem{Abbott_2009}
B.~P. Abbott and {The LIGO Scientific Collaboration}, \enquote{{LIGO}: the
  laser interferometer gravitational-wave observatory,}
  {\protect\JournalTitle{Reports on Progress in Physics}} \textbf{72}, 076901
  (2009).

\bibitem{DIAZORTIZ2022100968}
M.~{Diaz Ortiz}, J.~Gleason, H.~Grote, A.~Hallal, M.~Hartman, H.~Hollis, K.-S.
  Isleif, A.~James, K.~Karan, T.~Kozlowski, A.~Lindner, G.~Messineo,
  G.~Mueller, J.~Põld, R.~Smith, A.~Spector, D.~Tanner, L.-W. Wei, and
  B.~Willke, \enquote{Design of the {ALPS II} optical system,}
  {\protect\JournalTitle{Physics of the Dark Universe}} \textbf{35}, 100968
  (2022).

\bibitem{HALLAL2022100914}
A.~Hallal, G.~Messineo, M.~D. Ortiz, J.~Gleason, H.~Hollis, D.~Tanner,
  G.~Mueller, and A.~Spector, \enquote{The heterodyne sensing system for the
  {ALPS II} search for sub-ev weakly interacting particles,}
  {\protect\JournalTitle{Physics of the Dark Universe}} \textbf{35}, 100914
  (2022).

\bibitem{C8RA04491K}
N.~Kuhar, S.~Sil, T.~Verma, and S.~Umapathy, \enquote{Challenges in application
  of raman spectroscopy to biology and materials,} {\protect\JournalTitle{RSC
  Adv.}} \textbf{8}, 25888--25908 (2018).

\bibitem{molecules27196752}
E.~I. Nagaev, I.~V. Baimler, A.~S. Baryshev, M.~E. Astashev, and S.~V. Gudkov,
  \enquote{Effect of laser-induced optical breakdown on the structure of bsa
  molecules in aqueous solutions: An optical study,}
  {\protect\JournalTitle{Molecules}} \textbf{27} (2022).

\bibitem{Hercher:67}
M.~Hercher, \enquote{An analysis of saturable absorbers,}
  {\protect\JournalTitle{Appl. Opt.}} \textbf{6}, 947--954 (1967).

\bibitem{MANILOFF1995173}
E.~S. Maniloff, F.~R. Graf, H.~Gygax, S.~B. Altner, S.~Bernet, A.~Renn, and
  U.~P. Wild, \enquote{Power broadening of the spectral hole width in an
  optically thick sample,} {\protect\JournalTitle{Chemical Physics}}
  \textbf{193}, 173--180 (1995).

\bibitem{Berger:16}
P.~Berger, Y.~Attal, M.~Schwarz, S.~Molin, A.~Louchet-Chauvet,
  T.~Chaneli\`{e}re, J.-L. {Le Gou\"{e}t}, D.~Dolfi, and L.~Morvan,
  \enquote{{RF} spectrum analyzer for pulsed signals: Ultra-wide instantaneous
  bandwidth, high sensitivity, and high time-resolution,}
  {\protect\JournalTitle{J. Lightwave Technol.}} \textbf{34}, 4658--4663
  (2016).

\bibitem{Venet:18}
C.~Venet, M.~Bocoum, J.-B. Laudereau, T.~Chaneliere, F.~Ramaz, and
  A.~Louchet-Chauvet, \enquote{Ultrasound-modulated optical tomography in
  scattering media: flux filtering based on persistent spectral hole burning in
  the optical diagnosis window,} {\protect\JournalTitle{Opt. Lett.}}
  \textbf{43}, 3993--3996 (2018).

\bibitem{NILSSON2005393}
M.~Nilsson and S.~Kröll, \enquote{Solid state quantum memory using complete
  absorption and re-emission of photons by tailored and externally controlled
  inhomogeneous absorption profiles,} {\protect\JournalTitle{Optics
  Communications}} \textbf{247}, 393--403 (2005).

\bibitem{Bussieres2014}
F.~Bussi{\`e}res, C.~Clausen, A.~Tiranov, B.~Korzh, V.~B. Verma, S.~W. Nam,
  F.~Marsili, A.~Ferrier, P.~Goldner, H.~Herrmann, C.~Silberhorn, W.~Sohler,
  M.~Afzelius, and N.~Gisin, \enquote{Quantum teleportation from a
  telecom-wavelength photon to a solid-state quantum memory,}
  {\protect\JournalTitle{Nature Photonics}} \textbf{8}, 775--778 (2014).

\bibitem{walther2015:PhysRevA.92.022319}
A.~Walther, L.~Rippe, Y.~Yan, J.~Karlsson, D.~Serrano, A.~N. Nilsson,
  S.~Bengtsson, and S.~Kr\"oll, \enquote{High-fidelity readout scheme for
  rare-earth solid-state quantum computing,} {\protect\JournalTitle{Phys. Rev.
  A}} \textbf{92}, 022319 (2015).

\bibitem{Maring2017}
N.~Maring, P.~Farrera, K.~Kutluer, M.~Mazzera, G.~Heinze, and H.~de~Riedmatten,
  \enquote{Photonic quantum state transfer between a cold atomic gas and a
  crystal,} {\protect\JournalTitle{Nature}} \textbf{551}, 485--488 (2017).

\bibitem{Molmer2016:PhysRevA.94.053804}
K.~M\o{}lmer, Y.~{Le Coq}, and S.~Seidelin, \enquote{Dispersive coupling
  between light and a rare-earth-ion--doped mechanical resonator,}
  {\protect\JournalTitle{Phys. Rev. A}} \textbf{94}, 053804 (2016).

\bibitem{seidelin2019:PhysRevA.100.013828}
S.~Seidelin, Y.~{Le Coq}, and K.~M\o{}lmer, \enquote{Rapid cooling of a
  strain-coupled oscillator by an optical phase-shift measurement,}
  {\protect\JournalTitle{Phys. Rev. A}} \textbf{100}, 013828 (2019).

\bibitem{Julsgaard:07}
B.~Julsgaard, A.~Walther, S.~Kr\"{o}ll, and L.~Rippe, \enquote{Understanding
  laser stabilization using spectral hole burning,} {\protect\JournalTitle{Opt.
  Express}} \textbf{15}, 11444--11465 (2007).

\bibitem{PhysRevLett.107.223202}
Q.-F. Chen, A.~Troshyn, I.~Ernsting, S.~Kayser, S.~Vasilyev, A.~Nevsky, and
  S.~Schiller, \enquote{Spectrally narrow, long-term stable optical frequency
  reference based on a
  {${\mathrm{Eu}}^{3+}:{\mathrm{Y}}_{2}{\mathrm{SiO}}_{5}$} crystal at
  cryogenic temperature,} {\protect\JournalTitle{Phys. Rev. Lett.}}
  \textbf{107}, 223202 (2011).

\bibitem{Thorpe2011}
M.~J. Thorpe, L.~Rippe, T.~M. Fortier, M.~S. Kirchner, and T.~Rosenband,
  \enquote{Frequency stabilization to $6\times 10^{-16}$ via spectral-hole
  burning,} {\protect\JournalTitle{Nature Photonics}} \textbf{5}, 688--693
  (2011).

\bibitem{Thorpe_2013}
M.~J. Thorpe, D.~R. Leibrandt, and T.~Rosenband, \enquote{Shifts of optical
  frequency references based on spectral-hole burning in
  {${\mathrm{Eu}}^{3+}:{\mathrm{Y}}_{2}{\mathrm{SiO}}_{5}$},}
  {\protect\JournalTitle{New Journal of Physics}} \textbf{15}, 033006 (2013).

\bibitem{Galland2020:PhysRevApplied.13.044022}
N.~Galland, N.~Lu\ifmmode \check{c}\else \v{c}\fi{}i\ifmmode~\acute{c}\else
  \'{c}\fi{}, B.~Fang, S.~Zhang, R.~Le~Targat, A.~Ferrier, P.~Goldner,
  S.~Seidelin, and Y.~{Le Coq}, \enquote{Mechanical tunability of an
  ultranarrow spectral feature of a rare-earth-doped crystal via uniaxial
  stress,} {\protect\JournalTitle{Phys. Rev. Applied}} \textbf{13}, 044022
  (2020).

\bibitem{Zhang2020aPhysRevResearch.2.013306}
S.~Zhang, N.~Galland, N.~Lu\ifmmode \check{c}\else
  \v{c}\fi{}i\ifmmode~\acute{c}\else \'{c}\fi{}, R.~Le~Targat, A.~Ferrier,
  P.~Goldner, B.~Fang, Y.~{Le Coq}, and S.~Seidelin, \enquote{Inhomogeneous
  response of an ion ensemble from mechanical stress,}
  {\protect\JournalTitle{Phys. Rev. Research}} \textbf{2}, 013306 (2020).

\bibitem{Zhang2020:doi:10.1063/5.0025356}
S.~Zhang, N.~Lučić, N.~Galland, R.~Le~Targat, P.~Goldner, B.~Fang,
  S.~Seidelin, and Y.~{Le Coq}, \enquote{Precision measurements of
  electric-field-induced frequency displacements of an ultranarrow optical
  transition in ions in a solid,} {\protect\JournalTitle{Applied Physics
  Letters}} \textbf{117}, 221102 (2020).

\bibitem{Zhang2023:PhysRevA.107.013518}
S.~Zhang, S.~Seidelin, R.~Le~Targat, P.~Goldner, B.~Fang, and Y.~{Le Coq},
  \enquote{First-order thermal insensitivity of the frequency of a narrow
  spectral hole in a crystal,} {\protect\JournalTitle{Phys. Rev. A}}
  \textbf{107}, 013518 (2023).

\bibitem{philippe_frequency_2016}
C.~Philippe, R.~Le~Targat, D.~Holleville, M.~Lours, T.~Minh-Pham, J.~Hrabina,
  F.~Du~Burck, P.~Wolf, and O.~Acef, \enquote{Frequency tripled
  $1.5\,\mathrm{{\mu}m}$ telecom laser diode stabilized to iodine hyperfine
  line in the $10^{-15}$ range,} in \emph{2016 European Frequency and Time
  Forum (EFTF),}  (2016), pp. 1--3.

\bibitem{Nicholson2012:PhysRevLett.109.230801}
T.~L. Nicholson, M.~J. Martin, J.~R. Williams, B.~J. Bloom, M.~Bishof, M.~D.
  Swallows, S.~L. Campbell, and J.~Ye, \enquote{Comparison of two independent
  sr optical clocks with
  $1\mathbf{\ifmmode\times\else\texttimes\fi{}}{10}^{\ensuremath{-}17}$
  stability at ${10}^{3}\text{ }\text{ }\mathbf{s}$,}
  {\protect\JournalTitle{Phys. Rev. Lett.}} \textbf{109}, 230801 (2012).

\bibitem{Itano1993:PhysRevA.47.3554}
W.~M. Itano, J.~C. Bergquist, J.~J. Bollinger, J.~M. Gilligan, D.~J. Heinzen,
  F.~L. Moore, M.~G. Raizen, and D.~J. Wineland, \enquote{Quantum projection
  noise: Population fluctuations in two-level systems,}
  {\protect\JournalTitle{Phys. Rev. A}} \textbf{47}, 3554--3570 (1993).

\bibitem{N00NStates:doi:10.1126/science.1188172}
I.~Afek, O.~Ambar, and Y.~Silberberg, \enquote{High-{NOON} states by mixing
  quantum and classical light,} {\protect\JournalTitle{Science}} \textbf{328},
  879--881 (2010).

\bibitem{Drever1983}
R.~W.~P. Drever, J.~L. Hall, F.~V. Kowalski, J.~Hough, G.~M. Ford, A.~J.
  Munley, and H.~Ward, \enquote{Laser phase and frequency stabilization using
  an optical resonator,} {\protect\JournalTitle{Applied Physics B}}
  \textbf{31}, 97--105 (1983).

\bibitem{Black2001:10.1119/1.1286663}
E.~D. Black, \enquote{{An introduction to Pound–Drever–Hall laser frequency
  stabilization},} {\protect\JournalTitle{American Journal of Physics}}
  \textbf{69}, 79--87 (2001).

\bibitem{Riza1996:10.1063/1.1147199}
N.~A. Riza, \enquote{{Scanning heterodyne optical interferometers},}
  {\protect\JournalTitle{Review of Scientific Instruments}} \textbf{67},
  2466--2476 (1996).

\bibitem{GNURadio_PPC}
\enquote{{Polyphase Channelizer},}
  \url{https://wiki.gnuradio.org/index.php/Polyphase_Channelizer}. [Online;
  published (at time of access) 31 December 2021].

\bibitem{GNURadio_Filter_Design}
\enquote{{GNU Radio Manual and C++ API Reference: Filter Signal Processing
  Blocks},} \url{https://www.gnuradio.org/doc/doxygen/page_filter.html}.
  [Online reference manual version 3.9.4.0, as accessed on 07 June 2023].

\bibitem{McIntyre1966:1474241}
R.~McIntyre, \enquote{Multiplication noise in uniform avalanche diodes,}
  {\protect\JournalTitle{IEEE Transactions on Electron Devices}}
  \textbf{ED-13}, 164--168 (1966).

\bibitem{Jobez2016:PhysRevA.93.032327}
P.~Jobez, N.~Timoney, C.~Laplane, J.~Etesse, A.~Ferrier, P.~Goldner, N.~Gisin,
  and M.~Afzelius, \enquote{Towards highly multimode optical quantum memory for
  quantum repeaters,} {\protect\JournalTitle{Phys. Rev. A}} \textbf{93}, 032327
  (2016).

\bibitem{Multi-mode_Heterodyne_Laser_Interferometry_Realized_via_Software_Defined_Radio:doi:10.5281/zenodo.8390942}
X.~Lin, M.~Hartman, S.~Zhang, S.~Seidelin, B.~Fang, and Y.~{Le Coq},
  \enquote{Dataset for: Multi-mode heterodyne laser interferometry realized via
  software defined radio,} \url{http://dx.doi.org/10.5281/zenodo.8390942}
  (2023).

\end{thebibliography}

\end{document}